\documentclass{aastex}
\usepackage{spr-astr-addons}
\usepackage{url}
\usepackage{hyperref}
\hypersetup{colorlinks,citecolor=blue,linkcolor=blue,urlcolor=blue}

\RequirePackage{color}

\newcommand{\emaila}{carindam1@gmail.com}

\begin{document}

\title{Study of magnetic field geometry and extinction in Bok globule CB130}
\slugcomment{Not to appear in Nonlearned J., 45.}
\shorttitle{Photopolarimetric study of CB130 cloud}
\shortauthors{Chakraborty \& Das.}

\author{A. Chakraborty\altaffilmark{$\dag$}}
\and
\author{H. S. Das\altaffilmark{$\star$}}
\affil{Department of Physics, Assam University, Silchar 788011, India}
\email{\emaila}{$^{\dag}$carindam1@gmail.com}
\email{\emaila}{\\$^{\star}$himadri.sekhar.das@aus.ac.in}\\


\begin{abstract}
We trace the peripheral magnetic field structure of Bok globule CB130, by estimating the linear polarization of its field stars in the R band. The magnetic field orientation sampled by these stars, aligned on average among themselves and the polarization produced within the cloud has a different direction from that of Galactic plane with an offset of 53$^\circ$. The offset between minor axis and the mean magnetic field of CB130 is found to be 80$^\circ$. The estimated strength of the magnetic field in the plane-of-the-sky is $\sim$ 116$\pm$19 $\mu$G. We constructed the visual extinction map using Near Infrared Color Excess (NICE) method to see the dust distribution around CB130. Contours of Herschel\footnote{Herschel is an ESA space observatory with science instruments provided by European-led Principal Investigator consortia and with important participation from NASA} SPIRE 500$\mu$m  dust continuum emission maps of this cloud is over-plotted on the visual extinction map which shows the regions having higher optical extinction corresponds to higher densities of dust. Three distinct high dust density cores (named as C1, C2, and C3) are identified in the extinction map. It is observed that the cores C1 and C3 located close to two previously known cores CB130-1 and CB130-2 respectively. Estimate of visual extinction of some moderately obscured stars of CB130 are made utilizing near-infra red photometry. Its observed that there is a feeble dependence of polarization on extinction; and polarization efficiency (defined as $ \textit{p}$/A$ _{V}$) of the dust grains decreases with the increase in extinction.
\end{abstract}



\section{Introduction}

\begin{sloppypar}
Bok Globules, introduced by \cite{b3} are the regions of relatively small isolated and dense, rounded globules. These globules are believed to be the areas of potential star formation and hence, are the most suitable candidates for studying the direct interplay between protostellar collapse, fragmentation and magnetic fields \citep{laz97, henn08}. The magnetic field in the outer region of a molecular cloud is often mapped in the optical wavelength, whereas the inner area is mapped in infrared and sub-millimeter wavelength, by measuring the linear polarization of the background stars \citep{bvss, bgj, b22, b11, bafg, bwskn, bfag,  b12, chak14,bwd}. The polarization of the radiation observed through a dusty medium is partly plane polarized, from the aligned dust grains in the interstellar medium. These dust grains are generally lined up with their long axes perpendicular to the magnetic field \citep{laz97, wh05, henn08}.

\citet{chak14} studied three globules CB 56, CB 60 and CB 69 to map the magnetic field within the observed region of the cloud. They estimated the extinction of field stars from $BVR$ magnitudes and identified the probable location of each field star (background/foreground/within the cloud). The polarization efficiency of these clouds was also studied which showed a decrease in polarization efficiency with an increase in extinction along the observed line of sight. \citet{bwd} presented the polarimetric observations in the optical and near-infrared of the three Bok globules B 335, CB 68 and CB 54, which were combined with archival observations in the sub-millimeter and the optical. They traced the magnetic field structures of these globules over a range of $10^2 - 10^5$ AU, covering optically thin and optically thick regions. Recently, \citet{j15} studied near-IR polarimetry data of background stars shining through a selection of starless cores taken in the $K$ band, probing visual extinctions up to A$_{V} \sim$  48 (mag). They found that polarization efficiency decreases with increase in A$_{V}$ with a power law slope roughly $-0.5$. However, at greater optical depths they found no grain alignment. If one accepts
the theory of dust grain alignment via radiative torques, this lack of alignment at greater optical depths may be due to the absence of a radiation source \citep{alve14,ande15}.

 As pointed out by \cite{lada04}, there are two conventional methods viz. the star count and the Near Infrared Color Excess (NICE) method, used for measuring extinction of background stars in the observed line of sight. In the star count method, extinction values are measured, at the expense of angular resolution which in turn losses the details of structural information \citep{cam99}. However, multi-wavelength Near Infrared (NIR) color excess method can measure extinction value at much deeper optical depth with significantly improved angular resolution with smaller uncertainties \citep{kan05}. Various NIR extinction mapping techniques (NICE, NICER, NICEST) have been used in the recent years to study the detailed structure of the clouds (viz. the distribution of the dust, temperature, density and the stability of the core) \citep{lada04, lom01, lom05, lom09, row09, bock12}.

We have taken polarimetric observations of CB130 at the R-band, with an aim to measure the optical polarization of background field stars to map the magnetic fields within the globule. The values of visual extinction ($A_{V}$) for background field stars are estimated using E(J-K) method \citep{bwth}. In this work, we have made a combined study of optical polarization and extinction of background field stars of CB130. We have also created a visual extinction map of CB130, using the stellar color excess method as described by \citet{row09}, which is a generalized version of NICE mapping technique developed by \citet{lada04}. The rest of the paper is organized as follows. In section 2, we present a brief description of CB130. Observation details and data reduction techniques are discussed in section 3. Finally, a detail discussion of results obtained are given in section 4, and a set of conclusions based on our work is presented in section 5.

\section{Description of the target (CB130)}
CB130 (L507) is an elongated globule, situated in the Aquila Rift region (galactic coordinates  $20^{\circ}<l<40^{\circ}$ and $-6^{\circ}<b<+14^{\circ}$) \citep{dame85}.  \citet{b4} first cataloged CB130, along with 248 other small isolated molecular cloud. \citet{b23} have categorized CB130 as ``A'' type cloud, by comparing CO peak line, temperature and CO line width of 248 small molecular clouds. ``A'' type cloud comprises a maximum number of clouds i.e. 74$\%$ of 248 clouds, where gas temperatures are cold ($\sim$ 8.5 K) and have tiny turbulent gas motion. CB130 is found to be located at a distance of 250$\pm$50  parsec (pc) \citep{laun97, stair03}. \citet{lee99} detected three cores in CB 130 from south to north and named them as CB 130-1, CB 130-2 and CB 130-3 whose central coordinates are shown in Table \ref{tab1}. It is to be noted here that CB 130-1 represents the central core of the globule CB 130. \citet{vey07} detected two YSOs in CB 130-1 using a three-color (3.6 $\mu m$ as blue, 4.5 $\mu m$ as green, and 8.0 $\mu m$ as red) image of CB130-1 using $\emph{Spitzer}$ IRAC images and GO-2 program (cores2deeper). \citet{kim11} later named these two YSOs as CB130-1-IRS1 and CB130-1-IRS2, which are 15$''$ apart, corresponds to 4100 AU, further, they mapped CB130-1 region with CO (J =2$\rightarrow$1) at Caltech Submillimeter Observatory (CSO)\footnote{The Caltech Submillimeter Observatory is supported by the NSF.} and found no significant evidence of out-flowing gas. \cite{laun10} also detected that CB130 core contains two well resolved compact sources (named them as SMM1 and SMM2), and observed that SMM1 is brighter than SMM2. Further, SMM1 is associated with a faint NIR source, a very red star $\sim$ 3000 AU east of SMM1 \citep{laun10, laun13}.  Recently, \citet{laun13} detected previously known embedded heating sources like protostars, including the VeLLO (very low-luminosity object) in CB 130, based on the temperature and the Herschel maps. They didn't detect any new source other than the previously known warm, compact source in the globule.

\section[]{OBSERVATION}
\subsection{Polarimetry and instrumentation}
Polarimetric observation of Bok globule CB130  was  carried out at R-band with exposure time of 600 sec in six nights, namely 26th, 28th and 30th April, 2014  and  2nd, 3rd  and 4th  May, 2014, from the 2-m Cassegrainian focus telescope (focal ratio of f/10) of Girawali Observatory at Inter University Center for Astronomy and Astrophysics (IUCAA), Pune (IGO, Latitude: 19$^\circ $5$'$N, Longitude: + 73$^\circ $40$'$ E, Altitude = 1000 m), India.  The IUCCA Faint Object Spectrograph and Camera (IFOSC) is the main instrument attached with the telescope, which is equipped with an EEV 2K $\times$ 2K pixel$^2$ CCD camera and an imaging  polarimeter  of FOV $\sim$ 2 arcmin radius, to measure linear polarization in the wavelength band $350-850$ nm. This instrument also uses a half-wave plate (HWP) followed by Wollaston prism to observe two orthogonal polarization components that define a Stokes parameter. For further details please refer to \cite{chak14} and references therein. The average value of FWHM for the stellar images is $\sim$ 2 arcsec. The observations were made for four sub-regions to cover CB 130-1 core.

Description of the steps followed for data reduction, and calibration of results are same as discussed in \cite{chak14}.	Instrumental polarization is determined by observing an unpolarized standard star HD 115617, and its polarization value in R-band, is found to be $\textit{p}<$ 0.05$\%$ which is in good agreement with literature \citep{bsmf}. We have also observed a polarized standard star HD 154445 at R-band taken from \cite{hsu} (our results: $\textit{p}_{obs}$ = (3.48 $\pm$ 0.06)\%, $\theta_{obs}$ = 88.90 $\pm$ 0.87 degrees) to calibrate our result with zero-position angle. In table \ref{tab2}, we have presented the estimated values of linear polarization at R-band for 54 field stars of CB 130.

\begin{table*}
 \begin{center}
   \caption{Central coordinates of three cores of CB130 (L507) (galactic coordinates$^\textrm{a}$: \textit{l} = 26.62$^\circ $, \textit{b} = 6.65$ ^\circ$).}
  \begin{tabular}{cccccccc}

\hline
Serial  & Object  &  RA(J2000)  & Dec.(J2000)  & Position Angle  &References &  $\theta_{GP}$   \\
number &      ID  &  (h m s)    &  $ (^\circ $ $^{\prime}$  $ ^{\prime\prime}) $ & long axis ( $ ^\circ $)& &( $ ^\circ $)  \\

\hline

1 & CB130/CB130-1 &  18 16 15.9 & $-$02 33 01& 90   &a, b & 27 \\
 2 & CB130-2 &	18 16 14.1 & $-$02 23 23&  & b & \\
 3 & CB130-3 &	18 16 17.9 & $-$02 16 41&  & b & \\

\hline
\end{tabular}

			\tablenotetext{a} {\citet{b4}}.
			\tablenotetext{b} {\citet{lee99}}.
			\tablecomments {[$\theta_{GP}$] Position angle of the galactic plane at \textit{b} = 6.65$ ^\circ$}.

\label{tab1}
\end{center}
\end{table*}

\subsection{Near-infrared data}
The near-infrared, $J $(1.24 $\mu m$), $H $(1.66 $\mu m$), and $K $(2.16 $\mu m$) magnitude of the field stars of CB130 have been obtained from the 2MASS Point Source Catalog in regions of $25' \times 25'$ centered on the globule \citep{cutri03}. Only those stars are selected whose $JHK$ magnitude are of the highest quality flag in each of the three filters (Qflag = ``AAA") i.e. SNR $\geq$10.


\begin{figure*}
\vspace{45pt}
\begin{center}
\includegraphics[width=25pc, height=22pc]{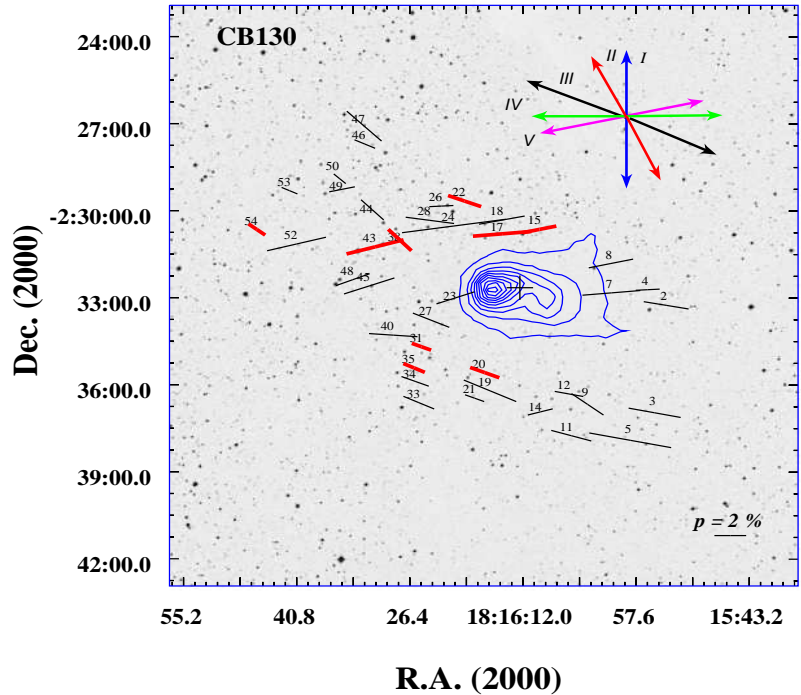}
\caption{The stellar polarization vectors and position angles are superimposed on a $\sim$ 20$' \times 20'$ R-band DSS image of the field containing CB130. We have plotted 39 stars whose, $\textit{p}/\in_{p} \geq3$. A vector with a polarization of 2$\%$ and 90$^\circ$ orientation w.r.t to the North is drawn for reference, the length of all the polarization vector is proportional to it.  The `+' symbol represents the globule center RA = 18h 16m 15.9s and Dec = $-$ 02d 33m 01s. Arrows I, II, III, IV, and V indicate the orientation of the minor axis with P.A. ($\sim$ 0 deg.), Galactic plane at b= 6.65$^\circ$ having P.A. ($\sim$ 27 deg.), avg. polarization P.A. ($\sim$ 80 deg.) of the observed field stars, the major axis of the cloud with P.A. ($\sim$ 90 deg.), and avg. polarization P.A. ($\sim$ 102 deg.) of some  stars  obtained from  \citet{hei00}, respectively in the observed plane of sky. We  have also over-plotted contours of Herschel SPIRE 500$\mu$m dust continuum emissions which range from 25 to 95, increasing in a step size of  10 mJy beam$^{-1}$.}
\label{Fig1}
\end{center}
\end{figure*}

\begin{table*}
\tabletypesize{\scriptsize}
\small
\caption{Estimated values of $p$ and $\theta$  of field stars in CB 130 at R-band along with  A$_{V}$ of some field stars with $A_{V}/(\Delta A_{V}) \geq 2$.}

\begin{tabular}{@{}ccccccccccccccc@{}}
\tableline
S/N   & ~RA(J2000)  & Dec.(J2000) &  $\textit{p}$ & ($\theta$ ) & $\textit{p}/\in_{p} \geq3$ & A$_{V}$ &$\textit{p}/A_{V}$$^\ddag$ & \\
   & (h m s)    & $ (^\circ $ $^{\prime}$  $ ^{\prime\prime} $) & $(\%)$ &  $(^\circ)$& (Y/N)$^{\odot}$& (mag)& ($\%$ mag$^{-1}$)&  \\
\tableline

1	&	18 15 55.2	&	$-$	02 37 19.6	&	5.49	$\pm$	2.01	&	85	$\pm$	10	&	N	&	\P	&	---	&	\\
2	&	18 15 56.1	&	$-$	02 33 23.2	&	2.87	$\pm$	0.50	&	80	$\pm$	22	&	Y	&	1.49	$\pm$	0.40	&	1.93	$\pm$	0.63	&	\\
3	&	18 15 57.6	&	$-$	02 37 05.5	&	3.34	$\pm$	0.53	&	79	$\pm$	08	&	Y	&	1.21	$\pm$	0.46	&	2.76	$\pm$	1.09	&	\\
4	&	18 15 58.4	&	$-$	02 32 50.9	&	1.51	$\pm$	0.37	&	93	$\pm$	07	&	Y	&   $\bigstar$ &	---			&	\\
5	&	18 16 00.7	&	$-$	02 38 02.4	&	5.25	$\pm$	0.92	&	79	$\pm$	05	&	Y	&	$\bigstar$			&	---			&	\\
6	&	18 16 01.4	&	$-$	02 30 41.6	&	3.63	$\pm$	2.35	&	85	$\pm$	19	&	N	&	$\bigstar$			&	---			&	\\
7	&	18 16 03.1	&	$-$	02 32 57.6	&	3.66	$\pm$	0.26	&	95	$\pm$	02	&	Y	&	2.48	$\pm$	0.49	&	1.15	$\pm$	0.33	&	\\
8	&	18 16 03.1	&	$-$	02 31 56.8	&	2.86	$\pm$	0.59	&	102	$\pm$	06	&	Y	&	1.71	$\pm$	0.41	&	1.46	$\pm$	0.59	&	\\
9	&	18 16 06.1	&	$-$	02 36 48.1	&	2.49	$\pm$	0.81	&	54	$\pm$	13	&	Y	&	3.13	$\pm$	0.43	&	0.84	$\pm$	0.3	&	\\
10	&	18 16 07.6	&	$-$	02 35 33.2	&	2.08	$\pm$	0.99	&	105	$\pm$	14	&	N	&	${\dag}$			&	---			&	\\
11	&	18 16 08.2	&	$-$	02 37 52.1	&	2.62	$\pm$	0.87	&	74	$\pm$	20	&	Y	&	2.3	$\pm$	0.46	&	0.79	$\pm$	0.46	&	\\
12	&	18 16 08.5	&	$-$	02 36 26.7	&	1.83	$\pm$	0.55	&	79	$\pm$	12	&	Y	&	2.18	$\pm$	0.46	&	1.45	$\pm$	0.99	&	\\
13	&	18 16 10.5	&	$-$	02 36 22.8	&	1.82	$\pm$	0.80	&	81	$\pm$	16	&	N	&	1.23	$\pm$	0.43	&	3.00	$\pm$	1.29	&	\\
14	&	18 16 12.1	&	$-$	02 37 03.9	&	1.62	$\pm$	0.52	&	105	$\pm$	09	&	Y	&	$\blacklozenge$			&	---			&	\\
15	&	18 16 12.2	&	$-$	02 30 46.3	&	2.17	$\pm$	0.06	&	102	$\pm$	01	&	Y	&	${\dag}$			&	---			&	\\
16	&	18 16 12.5	&	$-$	02 28 54.6	&	3.16	$\pm$	1.56	&	84	$\pm$	19	&	N	&	\P	&	---	&	\\
17	&	18 16 16.9	&	$-$	02 30 55.6	&	3.69	$\pm$	0.92	&	95	$\pm$	07	&	Y	&	\P	&	---	&	\\
18	&	18 16 16.9	&	$-$	02 30 27.3	&	2.93	$\pm$	0.81	&	101	$\pm$	08	&	Y	&	$\blacklozenge$			&	---			&	\\
19	&	18 16 18.5	&	$-$	02 36 20.3	&	3.62	$\pm$	1.20	&	66	$\pm$	12	&	Y	&	$\blacklozenge$			&	---			&	\\
20	&	18 16 19.2	&	$-$	02 35 42.6	&	1.97	$\pm$	0.34	&	69	$\pm$	05	&	Y	&	${\dag}$		&	---			&	\\
21	&	18 16 20.5	&	$-$	02 36 35.7	&	1.29	$\pm$	0.19	&	69	$\pm$	06	&	Y	& 	$\blacklozenge$			&	---			&	\\
22	&	18 16 21.7	&	$-$	02 29 47.9	&	2.22	$\pm$	0.60	&	70	$\pm$	08	&	Y	&	${\dag}$			&	---			&	\\
23	&	18 16 22.9	&	$-$	02 33 08.1	&	2.53	$\pm$	0.31	&	109	$\pm$	04	&	Y	&	$\bigstar$			&	---			&	\\
24	&	18 16 23.1	&	$-$	02 30 39.4	&	6.60	$\pm$	1.36	&	98	$\pm$	06	&	Y	&	$\blacklozenge$			&	---			&	\\
25	&	18 16 23.1	&	$-$	02 29 32.1	&	1.82	$\pm$	0.91	&	71	$\pm$	16	&	N	&	1.65	$\pm$	0.46	&	2.03	$\pm$	0.63	&	\\
26	&	18 16 24.7	&	$-$	02 29 58.3	&	1.52	$\pm$	0.50	&	93	$\pm$	10	&	Y	&	$\blacklozenge$		&	---			&	\\
27	&	18 16 26.0	&	$-$	02 33 54.3	&	2.49	$\pm$	0.82	&	67	$\pm$	17	&	Y	&	2.36	$\pm$	0.29	&	0.60	$\pm$	0.11	&	\\
28	&	18 16 26.2	&	$-$	02 30 27.8	&	3.08	$\pm$	0.24	&	82	$\pm$	02	&	Y	&	$\bigstar$		&	---			&	\\
29	&	18 16 26.2	&	$-$	02 35 38.4	&	2.10	$\pm$	1.94	&	66	$\pm$	26	&	N	&	$\bigstar$			&	---			&	\\
30	&	18 16 26.4	&	$-$	02 35 28.9	&	2.07	$\pm$	1.22	&	66	$\pm$	17	&	N	&	${\dag}$			&	---			&	\\
31	&	18 16 27.2	&	$-$	02 34 49.2	&	1.30	$\pm$	0.23	&	69	$\pm$	05	&	Y	&	\P	&	---	&	\\
32	&	18 16 27.3	&	$-$	02 32 12.9	&	1.27	$\pm$	0.65	&	66	$\pm$	15	&	N	&	$\bigstar$		&	---			&	\\
33	&	18 16 27.6	&	$-$	02 36 44.9	&	2.12	$\pm$	0.12	&	66	$\pm$	02	&	Y	&	1.89$\pm$0.41	&	1.12$\pm$0.25	&	\\		
34	&	18 16 28.0	&	$-$	02 36 00.8	&	1.84	$\pm$	0.59	&	69	$\pm$	09	&	Y	&	$\bigstar$		&	---			&	\\
35	&	18 16 28.2	&	$-$	02 35 33.1	&	1.48	$\pm$	0.22	&	66	$\pm$	04	&	Y	&	$\blacklozenge$	&	---			&	\\
36	&	18 16 29.2	&	$-$	02 36 00.3	&	4.43	$\pm$	2.41	&	95	$\pm$	16	&	N	&	$\bigstar$			&	---			&	\\
37	&	18 16 29.9	&	$-$	02 36 50.4	&	2.44	$\pm$	1.83	&	66	$\pm$	21	&	N	&	$\bigstar$		&	---			&	\\
38	&	18 16 29.1	&	$-$	02 31 08.6	&	2.07	$\pm$	0.12	&	45	$\pm$	02	&	Y	&	\P	&	---	&	\\
39	&	18 16 30.8	&	$-$	02 37 24.6	&	3.54	$\pm$	1.89	&	86	$\pm$	15	&	N	&	$\bigstar$		&	---			&	\\
40	&	18 16 30.8	&	$-$	02 34 25.3	&	3.06	$\pm$	0.84	&	86	$\pm$	23	&	Y	&	$\blacklozenge$		&	---			&	\\
41	&	18 16 31.6	&	$-$	02 36 55.9	&	2.54	$\pm$	1.25	&	65	$\pm$	15	&	N	&	 1.30 $\pm$	0.43	&	1.95$\pm$1.16	&	\\
42	&	18 16 31.7	&	$-$	02 37 33.8	&	1.51	$\pm$	1.23	&	57	$\pm$	23	&	N	&	${\dag}$			&	---			&	\\
43	&	18 16 33.1	&	$-$	02 31 22.8	&	3.72	$\pm$	0.37	&	105	$\pm$	03	&	Y	&	${\dag}$			&	---			&	\\
44	&	18 16 33.5	&	$-$	02 30 06.2	&	1.98	$\pm$	0.09	&	46	$\pm$	01	&	Y	&	$\blacklozenge$		&	---			&	\\
45	&	18 16 33.9	&	$-$	02 32 43.6	&	3.36	$\pm$	0.44	&	109	$\pm$	04	&	Y	&	$\bigstar$	 & ---     & 	\\					
46	&	18 16 34.4	&	$-$	02 27 50.1	&	1.42	$\pm$	0.20	&	66	$\pm$	13	&	Y	&	1.30	$\pm$	0.49	&	1.09	$\pm$	0.44	&	\\
47	&	18 16 34.5	&	$-$	02 27 13.0	&	2.99	$\pm$	0.16	&	47	$\pm$	02	&	Y	&	$\blacklozenge$				&	---			&	 \\
48	&	18 16 35.1	&	$-$	02 32 30.6	&	2.27	$\pm$	0.20	&	111	$\pm$	03	&	Y	&	1.12	$\pm$	0.46	&	2.03	$\pm$	0.76	&	\\
49	&	18 16 37.3	&	$-$	02 29 23.8	&	1.65	$\pm$	0.49	&	102	$\pm$	08	&	Y	&	$\blacklozenge$			&	---			&	\\
50	&	18 16 37.6	&	$-$	02 29 01.1	&	1.00	$\pm$	0.19	&	49	$\pm$	05	&	Y	&	0.83	$\pm$	0.41	&	1.21	$\pm$	0.64	&	\\
51	&	18 16 41.7	&	$-$	02 32 01.3	&	1.05	$\pm$	0.51	&	118	$\pm$	14	&	N	&	1.11	$\pm$	0.41	&	0.95	$\pm$	0.59	&	\\
52	&	18 16 43.1	&	$-$	02 31 16.9	&	3.82	$\pm$	0.90	&	104	$\pm$	07	&	Y	&	$\bigstar$			&	---			&	\\
53	&	18 16 44.0	&	$-$	02 29 26.6	&	1.09	$\pm$	0.26	&	66	$\pm$	18	&	Y	&	$\blacklozenge$				&	---			&	 \\
54	&	18 16 48.2	&	$-$	02 30 46.1	&	1.30	$\pm$	0.20	&	54	$\pm$	04	&	Y	&	${\dag}$		&	---			&	\\

\tableline
\end{tabular}

\label{tab2}

\tablecomments{\tiny{$^{\odot}$ Y (for Yes) and N (for No); $^{\ddag}$~ Polarization efficiency; $\P$  denotes those stars, which touches the intrinsic color curves (refer to Figure \ref{Fig2});  $\bigstar$~ represents those stars whose photometric errors in $J, H$ and $K$ are greater than 0.03 mag); ${\dag}$~  those stars having $A_{V}/(\Delta A_{V}) < 2$; $\blacklozenge$ stars whose extinction value cannot be determined using $E(J-K)$ method, as they can't be traced back to the intrinsic color curves.}}
\end{table*}


\section{RESULTS}

\subsection{Polarization Map}
In figure \ref{Fig1}, we have plotted the polarization vectors (also termed as pseudo vectors\footnote{These polarization measurements are not true vectors because they have an 180 degree ambiguity}) of 39 stars (having $p/\in_{p}\geq 3$, $\in_{p}$ is the error in polarization value), which represent the projection of magnetic field on the observed plane of the sky. The polarization vectors are superimposed on DSS (Digital Sky Survey) image of CB130, taking observed field stars as their centers. The length of the polarization vector is proportional to the magnitude of polarization $\textit{p}$ (in percent), and it is aligned in agreement with position angle $\theta$ (in degree),  measured w.r.t north increasing eastward. All the vectors are scaled to a reference line drawn at the bottom right corner with $\textit{p}= 2\%$  and $\theta=90^{\circ}$.

\begin{figure*}
\begin{center}
\vspace{-2.3cm}
\includegraphics[width=20pc, height=30pc, angle= -90]{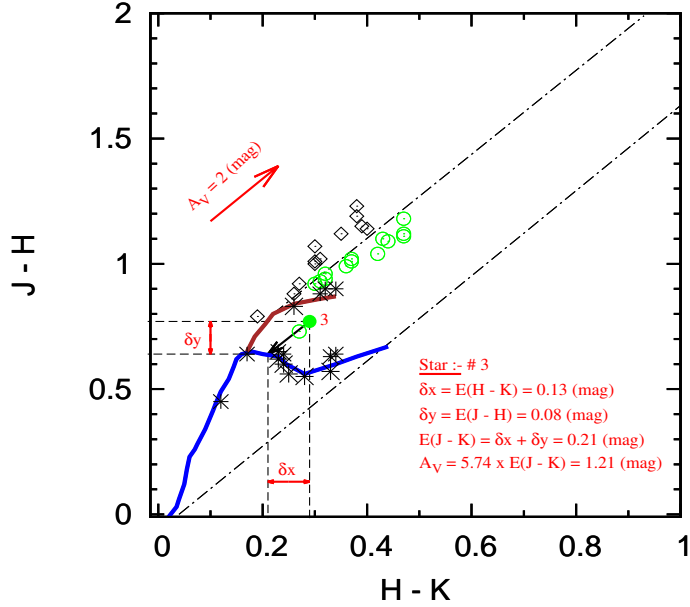}
\caption{$(J-H)$ vs $(H-K)$ color-color diagram for 42 field stars of CB 130 (whose photometric errors in $J$, $H$ and $K$  are $\leq$0.03 mag), listed in table \ref{tab2}. Blue and brown colored curves represent the intrinsic colors for normal dereddened stars of main-sequence (classes V) and the red giant branch (classes III), respectively\protect\citep{stair09}. An arrow in the upper left corner of the diagram is drawn with a slope of 1.70 \protect\citep{dr03}, represents the sample reddening vector of visual extinction ($A_{V}$) = 2 (in mag). Out of these 42 stars we could estimate the value of $A_{V}$ for 29 field stars  which can be traced back to the intrinsic color curve along the standard reddening vector (green open circles  represent 16 field stars whose extinction values are listed in table \ref{tab2} and their $A_{V}/\Delta A_{V} \geq 2$, while the black asterisk represent 13 stars, which either touches intrinsic color curve or those having $A_{V}/\Delta A_{V} < 2$). Thirteen open black squares represent those stars, which can't be traced back to the intrinsic color curves. The two diagonal dash-dotted lines are drawn parallel to the reddening vector and thus, represent the upper and lower bounds of the area in which reddened field stars with normal intrinsic color are expected to be distributed. To explain the method stated in this paper for estimating the reddening of a star, a black arrow is drawn for the star (\#3) (as an example, marked by filled circle) parallel to the sample reddening vector from the observed color to the intrinsic color-color curve. The estimated values of $E(J-H)$, $E(H-K)$, $E(J-K)$, and hence $A_{V}$ are also shown in the figure (please see the text for details). }

\label{Fig2}
\end{center}
\end{figure*}

\subsection{Estimation of  visual extinction ($A_{V}$) from NIR photometry}
\subsubsection{Estimation of $A_{V}$ from $E(J-K)$}

Dust column density in a given line of sight is often expressed in terms of the extinction ($A_{V}$) that it would produce in the V photometric band. For stars lacking spectral classifications, researchers have used some indirect methods, based on estimating the color excess of a star, to determine the value of its $A_V$ \citep{ryd1976, Tex99, ber01, bwth, mlb10, chak14}. These methods exploit the fact that the extinction law in the interstellar medium is roughly constant over many lines of sight in the NIR \citep{car89}. Thus, even if the actual spectral type is unknown, an accurate estimate of the extinction can be done by employing this method. To determine the values of visual extinction for the observed field stars of CB 130, we have used the $E (J - K)$ method, which is extensively described in \cite{bwth, she08}.

 The value of visual extinction ($A_V$) to an object, can be determined from the $E(J-K)$ of an object, using the relation:
	\begin{equation}
	 A_V=r_{1} \times E(J-K)
	\label{eq1}	
	 \end{equation}


The value of $r_{1}$ for diffuse interstellar medium is given by 5.74 \citep{Tex99}. Figure \ref{Fig2} represents a $(J-H)$ vs $(H-K)$ color-color diagram for 42 fields stars\footnote{Here we have considered only those sources, whose uncertainties in J, H, and K filters are $\leq$0.03 mag.} of CB130, where the blue and brown colored curves represent the intrinsic colors for normal dereddened stars of main-sequence (classes V) and red giant branch (classes III), respectively \citep{stair09}. The arrow in the upper left corner of the diagram represents a sample reddening vector with a slope of 1.70 (defined by $E (J - H) / E (H - K)$) \citep{dr03}, having visual extinction ($A_{V}$) = 2 (mag). The value of $E(J-K)$ of a star is given by $E (J - K)= E(J-H) + E(H-K)$, where $E (J-H)$ = $[(J-H)_{obs} - (J-H)_{int}]$ and $E (H-K)$ = $[(H-K)_{obs} - (H-K)_{int}]$.  Our basic idea is to estimate the values of $E (J - K)$ for those field stars, whose reddening vectors can be traced back to the intrinsic color curve along the standard reddening vector. The value of these color excess is estimated in two steps first by extrapolating along the appropriate reddening vector onto intrinsic color lines, and then by projecting the reddening vector drawn onto the $(J-H)$ and $(H-K)$ axis, which in turn gives the magnitude of $E(J-H)$ and $E(H-K)$ respectively for a particular star. This method of estimating $A_{V}$, generally provides unambiguous results, as the bright background stars included in our sample are either late-type $(K, M)$ giants, which deredden onto the upper branch, or main-sequence stars earlier than $K0$; red dwarfs distant enough to be background to the cloud are predicted to be too dim at $2.16\mu m$ band to be selected in our sample. For further details please see \citep{bwth, she08}.

Green open circles in Figure \ref{Fig2} represent 16 field stars whose extinction values are listed in table \ref{tab2} and their $A_{V}/(\Delta A_{V}) \geq 2$, while the black asterisk represent 13 stars which touches the intrinsic color lines or those having $A_{V}/\Delta A_{V} < 2$ (where $\Delta A_{V}$ represents error in $A_{V}$). It is to be noted here that we considered those stars to have negligible extinction, which touches intrinsic color curves. However using above mentioned technique we couldn't estimate the value of  extinction values for these stars with greater certainty, thus we haven't considered them for our further analysis. Thirteen open black squares represent those stars, which can't be traced back to the intrinsic color curves. The average value of $A_{V}$ for 16 stars is ($A_{V})_{avg}$= 1.71 (mag) having a standard deviation of ($\sigma_{A_{V}}$) = 0.64 (mag).

\subsubsection{Visual extinction map}
The degree of linear polarization depends both on the properties of dust grains and the environment in which they exist. Therefore to understand the variation of the polarization with extinction ($A_{V}$), we have created a visual extinction map for CB130, using the Near-Infrared Color Excess (NICE) method as described by \citep{row09}. In this technique, the median color map has to be created, which can be further converted to the color excess map. The infrared color excess is directly related to the visual extinction via the extinction law \citep{row09}:		
 \begin{equation}
	 A_V=\frac{5.689}{2}(A_{H,<J-H>}+A_{H,<H-K>})
	\label{eq2}	
 \end{equation}
where, \\
\\
$A_{H,<J-H>}=\frac{<J-H>}{(\frac{\lambda_{H}}{\lambda_{J}})^{\beta}-1}$
or
$A_{H,<H-K>}=\frac{<H-K>}{1-(\frac{\lambda_{K}}{\lambda_{H}})^{-\beta}}$
\\

\begin{figure*}
\begin{center}
\vspace{-1.4 cm}
\includegraphics[width=32pc, height=22pc]{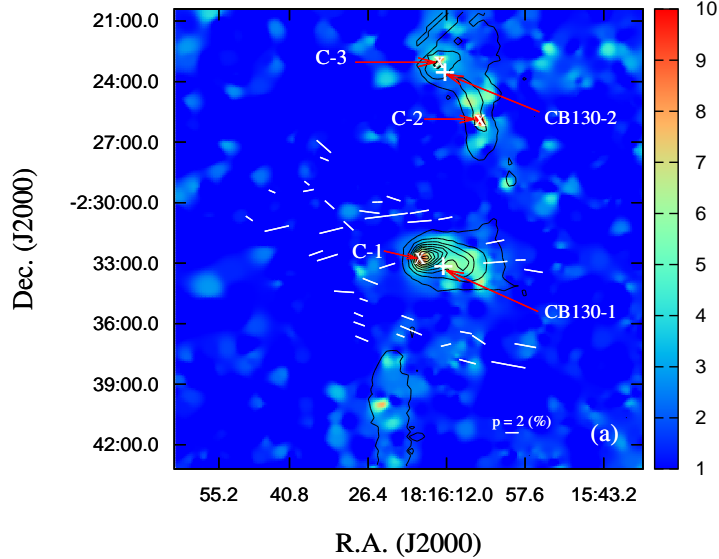}

\caption{Polarization vectors overlaid on visual extinction ($A_{V}$) map of CB 130 (map includes both CB130-1 and CB130-2 core) constructed using NICE method having field of view (FOV) $\sim 25' \times 25'$, dimension of a pixel $= 10 '' \times 10''$ (spatial resolution $= 34''$). Here we have plotted 39 stars whose $p/\in_{p} \geq3$. Polarization vectors are scaled with a reference vector overlaid at the bottom right corner with $p = $2$ \%$ and $\theta$ = 90$^{\circ}$. We have used two `+' symbols to represent two cores of CB 130: CB 130-1 core located at RA = 18h 16m 15.9s and Dec. = $-$ 02d 33m 01s (also the center of the cloud), and CB 130-2 core located at  RA = 18h 16m 14.1s and Dec.= $-$ 02d 23m 23s. Three distinct high dust density cores could be seen in the extinction map: one close to the center of the cloud marked by `C-1' and the other two in more northern part marked by `C-2' and `C-3' which are represented by the symbol `$\times$'. Contours correspond to Herschel SPIRE 500$\mu$m dust continuum emissions which range from 25 to 95, increasing a step size of 10 mJy beam$^{-1}$.}
\label{Fig3}
\end{center}
\end{figure*}

In deriving the extinction from equation \ref{eq2}, we adopted the coefficient $\beta$=1.7 \citep{dr03}. For further details of color-excess mapping technique and discussions, please see \citep{row09} and the references therein.

In Figure \ref{Fig3} (a), we have overlaid the polarization vectors of 39 stars (whose $p/\in_{p} \geq 3$) on the visual extinction map of CB130,  which represents the spatial variation of polarization with extinction in the observed plane of the sky. The field of view of visual extinction map is $\sim 25' \times 25'$, having dimension of each pixel $= 10 '' \times 10''$ (spatial resolution $= 34''$). We also have over-plotted contours,  corresponds to Herschel SPIRE 500$\micron$ dust continuum emission map (downloaded from Herschel Science Archive),  over the visual extinction map of CB130. These contours range from 25 to 95, increasing a step size of 10 mJy beams$^{-1}$.
 Three distinct high dust density cores could be seen in the extinction map: one close to the center of the cloud marked by `C-1' (J2000: RA= 18h 16m 17s, Dec.= $-$02d 32m 43s) and the other two in the northern part marked by `C-2' (J2000: RA = 18h 16m 06s, Dec. = $-$02d 25m 47s) and `C-3' (J2000: RA = 18h 16m 14.4s, Dec. = $-$02d 22m 55.2s). We can identify that `C-1' and  `C-3' cores are located very close to two cores CB130-1 and CB130-2 (please see Table 1) which were detected by \cite{lee99}. It is thus clearly observed from the figure that, the visual extinction map matches well with the dust continuum emission i.e. the regions in the map having higher visual extinction correspond to higher densities of dust. Further we found that the core $C1$ is identical to CB 130-1-IRS2 [\cite{kim11}: RA (2000) = 18h 16m 17.4s, Dec (2000) = $-$02d 32m 41.1s], which is also one of the YSO detected by \citet{vey07} in CB 130-1. \citet{laun10} and \cite{kim11} found that CB130-1-IRS1 is younger and more embedded than CB 130-1-IRS2. In our extinction map, we could detect CB130-1-IRS2, but not CB130-1-IRS1. It may be due to the gas density is high in that region as compared to dust density. Further, we couldn't associate $C2$ to any previously known sources in the literature, as we didn't find enough studies done around that region. So it will be very interesting to look into the part for detail understanding.

\begin{figure*}
\begin{center}
\vspace{-2 cm}
\hspace{-2 cm}
\includegraphics[width=30pc, height=20pc]{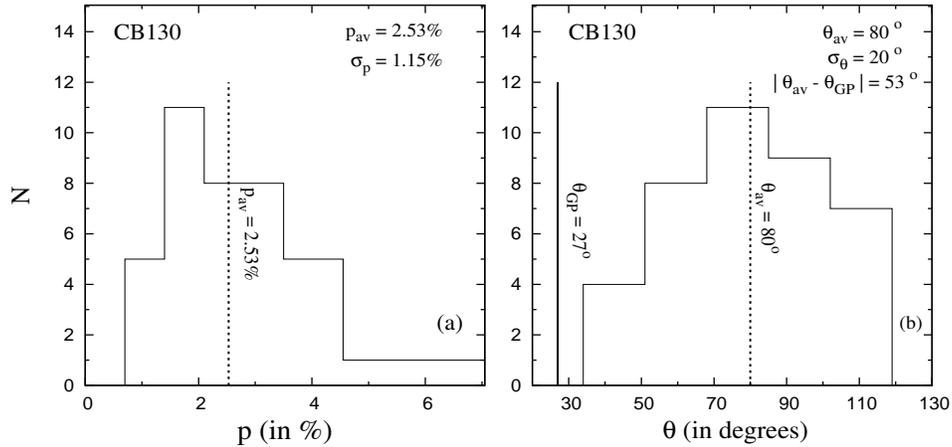}

\caption{ Histogram shows the distribution of the degree of linear polarization and polarization position angle vs. $N$ (Number of star), for 39 stars of CB130 having $\textit{p}/\in_{p} \geq3$. Dotted line in figure~\ref{Fig4} (a) \& (b) represents the position of the mean value of polarization vectors and position angle respectively for the observed field stars. A solid line is drawn in \ref{Fig4} (b) to indicate position angle ($\theta_{G.P}=27^{\circ}$) of the Galactic plane at b = +6.65$^\circ$.}
\label{Fig4}
\end{center}
\end{figure*}

\subsection{Geometry of magnetic field in CB130}
Figure \ref{Fig4} represents the distribution of polarization and position angle for the observed field stars (No. of stars = 39, those having $p/\in_{p}\geq 3$)  of CB130, in R band. The mean value of polarization of those stars is $p_{avg} = 2.53 \%$ with a standard deviation $\sigma_{p}$=1.15$\%$, also the mean value of the position angle is $\theta_{avg}$=80$^{\circ}$ which represents orientation of peripheral magnetic field with a standard deviation $\sigma_{\theta}$ = 20$^{\circ}$.
	
	It can be observed from the figure \ref{Fig1} that the polarization vectors of most of the stars are well aligned along a common direction in the observed plane of the sky. This better alignment of polarization vectors among themselves is entirely expected, as  CB 130 is  ``A"-type cloud (as mentioned in section 2), which have less dynamical activity and turbulence. The solid arrow (II) in the figure \ref{Fig1} is drawn parallel to the Galactic plane to represent the relative orientation of Galactic plane in the observed plane of the sky. The position angle of the galactic plane at the latitude of the cloud is found to be $\theta_{G.P}=27^{\circ}$. We found the angular offset between the position angle of the Galactic plane of the cloud and the mean value of observed polarization position angle is $|\theta_{G.P}-\theta_{avg}|= 53^{\circ}$, which indicates the polarization produced within the cloud has a different direction from that produced in the interstellar (IS) medium. In the IS medium, it is generally found that the polarization is mostly aligned along the direction of the Galactic magnetic field (coinciding with the direction of Galactic plane).
	
	The magnetic field geometry of a molecular cloud is believed to play a crucial role in contraction and subsequent star formation, as it provides the necessary support (in addition to thermal and turbulent pressure) to a molecular cloud that would otherwise collapse under its weight  (\cite{mou1991} and the other references therein).    Models of magnetically dominated star formation predict that the magnetic field should lie along the minor axis of the star-forming cloud \citep{mou1991,li98}, and it is expected that the cloud tends to contract first in a direction parallel to the magnetic field and then in quasi-statically perpendicular to the field orientation \citep{li98,bwskn}.
    The position angle (P.A.) of the major axis of CB130 is $\theta_{maj} = 90^{\circ}$ \citep{b4}, and that of minor axis is $\theta_{min}= 0^{\circ}$, indicating the observed magnetic field in the periphery of CB130 is found to be almost aligned with the major axis ($|\theta_{maj}-\theta_{avg}|= 10^{\circ}$), while the offset of the former with the minor axis of the cloud is ($|\theta_{min}-\theta_{avg}|= 80^{\circ}$). This is inconsistent with the magnetically dominated star formation models. A similar type of inconsistency between the minor axis and the peripheral magnetic field was seen by \cite{esw13} for L1570 and for IRAM 04191, L1521F, L673-7, L1014 by \cite{soam15}. However, it is also observed that the orientation of magnetic field structure in dense and heavily obscured region is not always parallel with the peripheral magnetic field. And absence of sub-millimeter data for CB130 have restricted our study to the low-density region only.

Further, to compare the relative orientation of peripheral magnetic field with the magnetic filed in the inter-cloud regions, we have obtained stellar polarization data of stars from \cite{hei00} within a circular area of radius 4$^{\circ}$ about the central coordinates of CB130.  We have found that the average value of position angle of those stars (No. of stars 4, found in 4$^{\circ}$ radius) is $\theta'_{avg} = 102^{\circ}$. We have also estimated the average value of polarization and visual extinction of stars located in 4$^{\circ}$ radius which are given by $p'_{avg}= 0.21\%, (A_{V})'_{avg}$= 0.23 (in mag) (taking R$_{V}$=3.1). Thus, it could be seen from above discussion that the magnetic field in the periphery of CB 130 has an offset of $|\theta'_{avg}-\theta_{avg}|= 22^{\circ}$ with the magnetic field of inter-cloud regions in the observed plane of the sky.

    It can also be seen from figure \ref{Fig1} and \ref{Fig3}, that the contours (Herschel SPIRE 500 $\mu$m ) overlaid on the extinction map at the center of the cloud CB130, are roughly elongated along east-west direction, and the magnetic field geometry of the cloud seems to follow this large-scale structure, as most of the polarization vectors overlaid, are well aligned with the elongation of these contours.

\subsubsection{Measuring magnetic field strength of CB130}

The magnetic field strength ($B_{pos}$) in the observed plane of the sky may be estimated by a modified version of the classical method proposed by \citep{cf53}, which generally provides a good estimate of the magnetic field strength in the observed plane-of sky, provided the dispersion in polarization angles is  $<25^{\circ}$\citep{ost01}. This technique assumes that the magnetic field is frozen into the gas, and that turbulence leads to isotropic fluctuation of the magnetic field around the mean field direction.

\begin{equation}
	B_{pos} = \sqrt{\frac{4\pi}{3}\bar{\rho}} ~\frac{\nu_{turb}}{\sigma_{\theta}}
\end{equation}
		
where $\bar{\rho}$ (in g $cm^{-3}$) and $\nu_{turb}$ (in cm $s^{-1}$) denote the density and $rms$ turbulence velocity of the gas respectively, while $\sigma_{\theta}$ denotes the standard deviation of the polarization position angles in radians.
Further $\bar{\rho}$= 1.36 $n_{H_{2}}M_{H_{2}}$;  where $M_{H_{2}}$ = 2.0158 amu = 2.0158 $\times$ 1.66$\times 10^{-24}$g is the mass of a H$_{2}$ molecule and $\nu_{turb}=\Delta\nu_{FWHM}/2.35$.

\begin{table*}
 \begin{center}
    \caption{Mean extinction, gas densities, gas velocities, and magnetic field strengths traced in the peripheral region of CB130.}
  \begin{tabular}{cccccccc}

\hline
Region & $\left\langle A_{V}\right\rangle$ & $n_{H_{2}}$& $\bar{\rho}$&$\Delta\nu_{FWHM}$  & $\nu_{turb}$  & $\sigma_{\theta}$ &  $B$   \\
        & (in mag)&$cm^{-3}$  & g$cm^{-3}$  & $km s^{-1}$  & $ km s^{-1}$  &  ($ ^\circ $)& ($\mu G$)  \\
\hline

 Peripheral & 1.74 &$2.65\times10^{3}$ &  1.21$\times10^{-20}$ & 4.2$^{\dag}$ & 1.79  & 20$\pm$3.20 & 116$\pm$19 \\

\hline
\end{tabular}

\tablecomments{$\dag$ \citet{lipp13}}.

\label{tab3}
\end{center}
\end{table*}

We have estimated the value of mean particle density $\left\langle n_{H_{2}}\right\rangle$ by using the relation

\begin{equation}
	\left\langle n_{H_{2}}\right\rangle = \left\langle A_{V}\right\rangle(\frac{N_{H_{2}}}{A_{V}})\frac{1}{l}
\end{equation}
assuming it to be a cylindrical filament.
	Here, $l $ (in cm) is the diameter of the cloud and $\left\langle A_{V}\right\rangle$ =1.74 is the mean extinction based on our study on section 4.2.1 and this value matches well with our extinction map. It is worthy to mention here that we have considered all the stars are lying behind the cloud.  Based on spectral energy distributions (SEDs), column density and dust temperature maps, \citet{laun13} estimated the mean radii of CB130 to be $R= 0.1 pc$. Also we have used standard gas-to-extinction ratio $(\frac{N_{H_{2}}}{A_{V}})= 0.94 \times10^{21} cm^{-2} mag^{-1}$ \citep{boh78} to estimate the mean molecular hydrogen column density. This ratio assumes that most of the hydrogen is in molecular form where $R_{V}$= 3.1. Further we assumed that CB130 is located at a distance of 250 pc (refer to section 2).

    Further uncertainty in our estimations of the magnetic field strength, arises from the errors in measuring, density of the core, turbulence velocity and standard deviation of position angles. However unavailability of uncertainties in both density of the core and turbulence velocity, have restricted us to determine the uncertainty in the magnetic field from the error associated with the standard deviation of polarization angles. We used, the standard error (S.E)\footnote{$S.E=\frac{\sigma_{\theta}}{\sqrt{N}}$ \citep{eve03}, here N represents sample size, for our case N=39} as a measure of precision in measuring standard deviation of polarization angle, which is found out to be 3.20$^\circ$. Using above data, we have estimated the mean magnetic field in the peripheral region of CB130 to be $\sim \textbf{116}\pm\textbf{19} \mu G$ which is listed in table \ref{tab3}.

 \subsection{Polarization efficiency}
\begin{figure*}
\begin{center}
\vspace{1 cm}
 \hspace{3cm}
\includegraphics[width=30pc, height=20pc]{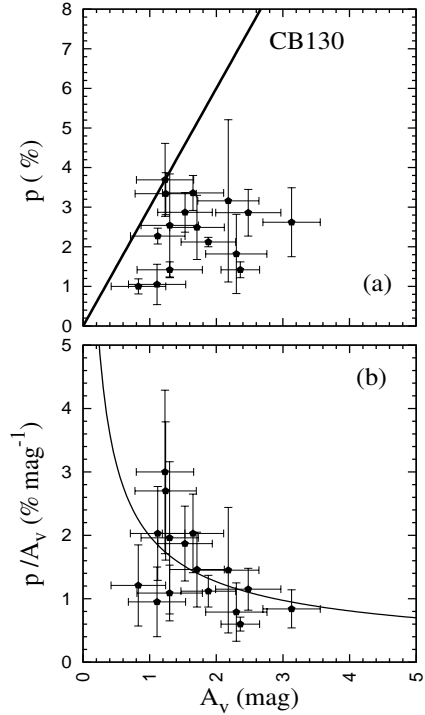}
\caption{(a) Upper panel is the plot of $A _{V} $ vs $ \textit{p} $ for 16 field stars of CB130. The solid line in the figure represents upper limit for polarization efficiency $ \textit{p} =3.0 A _{V} $\protect\cite{b150}. (b) Lower panel is the plot of $A _{V} $ vs $\textit{p} /A _{V}$ for the 16 background field stars of CB130. The solid line in the figure represents the  unweighted power-law fit to our sample $\textit{p} / A _{V}$ = $1.82A_{V}^{-0.62}$ (per cent mag$^{-1}$).}
\label{Fig5}
\end{center}
\end{figure*}

The non-spherical dust grains of the interstellar medium, are believed to be aligned with respect to the magnetic field by the interactions of the incident anisotropic radiation with the grains and the local magnetic field, which in turn produces differential extinction and hence polarize the background starlight. \cite{b22} found that the bending and distortions of magnetic field lines traced by optical polarization can be expected in regions where the accumulation of gas has occurred, or still occurring, with infall speeds comparable to or greater than the Alfven waves. The dependence of polarization (hence the grain alignment), on the visual extinction, is generally estimated by measuring \emph{polarization efficiency}, defined as the ratio of polarization produced for a given amount of extinction ($p/A _{V}$) \citep{b150}. Figure \ref{Fig5}(a), represents the variation of $A_{V}$ versus $\textit{p}$ for the field stars of CB130. It should be noted here that we have considered all the stars to be located in the background to the cloud. However tracing the magnetic field by estimating the linear polarization of background starlight in optical wavelength is limited to low extinction region only ($A_{V} \sim$ 1$-$5 (mag)) \citep{bvss} . It can also be observed that the field stars of CB130, lie below the line drawn by using the relation: $p(\%)= 3 A _{V}$ (mag) representing the optimal polarization efficiency of the grains due to selective extinction in the diffuse interstellar medium (ISM) \citep{b150}. \cite{b150} found $p/A _{V}$ = 3 (per cent mag$^{-1}$) for those regions where the magnetic field is disordered or not transverse to the line-of-sight, or the degree of grain alignment is lower. They also concluded that the interstellar dust grains must be sufficiently non-spherical and sufficiently aligned so that $p/A _{V}$ = 6 (per cent mag$^{-1}$). However the theoretical upper limit of the $p/A _{V}$ $\leq$ 14 (per cent mag$^{-1}$), for dust grains consisting of completely aligned infinite dielectric cylinders \citep{wh92}.

 Figure  \ref{Fig5}(b) displays the variation of $p/A_{V}$ vs. $A_{V}$ for the field stars of CB130, it indicates that polarization efficiency decreases with the increase in extinction, which suggests that there is a decline in the efficiency of grain alignment in the inner region as compared to the outer region of the dark cloud.  This observed variation in $p/A _{V}$ may arise due to various factors, such as the presence of turbulence in the magnetic field in the medium and/or various components of random/uniform magnetic fields oriented differently along the line of sight \citep{j92}. \citep{b10} also studied this phenomenon of a decrease in polarization with an increase in extinction for the Taurus dark cloud (TDC) by considering some factors (poor grain alignment, grain growth and/or changes in grain shape or composition). Recently \cite{hoang14, hoang15} showed that the radiative torques model can predict a fall-off of polarization efficiency with increasing extinction. This dependency of polarization efficiency on extinction can be well explained by using a power law of the form $p/A_V \propto A_V^{-\alpha}$. The solid line in the figure shows an unweighted power-law fit to our sample $p/A _{V}$ = $1.82 \pm 0.33 A _{V}^{-0.62\pm 0.22}$ (percent mag$^{-1}$). This result is consistent with the findings of \cite{b10, bwth, chak14} and \cite{barman15}. In \cite{chak14}, a similar dependence of polarization efficiency on extinction in the Bok globules CB56, CB60 and CB69 was reported.

\section{CONCLUSIONS}

\begin{enumerate}
	\item We have traced the local magnetic field structure of CB 130 in low-density region, by measuring the value of linear polarization in the optical wavelength for 30 field stars (whose $p/\in_{p}\geq 3$). Polarization map obtained from our study indicates that the polarization vectors of most of the stars are aligned in some common direction which in turn shows that magnetic field orientation sampled by these background stars appear to be aligned on average. The angular offset between the position angle of the Galactic plane of the cloud and the mean value of observed polarization position angle is 53$^{\circ}$, which indicates the polarization produced within the cloud has a direction different from that generated in the IS medium. Further, we have found that the minor axis is almost perpendicular to the mean peripheral magnetic field of CB130.

\item We have estimated the value of $\left\langle n_{H_{2}}\right\rangle$ = $2.65\times10^{3} gcm^{-3}$ for CB130, which is further used to calculate the mean magnetic field strength in the outer envelope which is given by $\sim 116\pm19 \mu G$.
	
	\item We have also presented the visual extinction map, on which polarization vectors of 39 field stars are overlaid, which represents the spatial variation of polarization with extinction in the observed plane of the sky of CB 130.  We have found that the variation of polarization with extinction is feeble in the low-density region of the cloud.    The extinction map is constructed using NICE method from NIR data, of dimension $\sim 25' \times 25'$. We have detected three distinct cores in CB130 and named them as $C1$, $C2$ and $C3$ from south to north. Cores $C1$ and $C3$ are found to be located very close to the central coordinates of CB 130-1 and CB 130-2. Further, we have observed that $C3$ is identical with CB130-1-IRS2 detected by \cite{kim11}. We couldn't associate $C2$ to any previously known sources in the literature, as we didn't find enough studies done around that region.

	\item Contours of Herschel SPIRE 500$\micron$ map, are overlaid on the extinction map at the center of the cloud CB130. It is observed that the contours are roughly elongated along north-east to southwest direction, and the magnetic field geometry of the cloud in the periphery seems to follow this large-scale structure, as most of the polarization vectors overlaid are well aligned with the elongation of these contours.
	
\item Visual extinction ($A_{V}$) for 15 field stars of CB 130 has been estimated employing $E(J-K)$ method.  We further found that the background starlight of CB 130 shows a tendency to decrease in polarization efficiency ($p/A_{V}$) with the increase in extinction. This suggests that there is a decline in the efficiency of grain alignment in the inner region as compared to the outer region of the dark cloud. Our results are in agreement with the findings of other clouds by different investigators.

\end{enumerate}

\end{sloppypar}

\acknowledgments
We gratefully acknowledge IUCAA, Pune for making telescope time available. We are also grateful to Dr. V. Mohan of IUCAA for helping us in probing CB 130 in polarimetric mode. The anonymous reviewer of this paper is highly acknowledged for his constructive comments which definitely contributed to improving the quality of the paper. We also acknowledge the use of the VizieR database of astronomical catalogues namely Two Micron All Sky Survey (2MASS), which is a joint project of the University of Massachusetts and the Infrared Processing and Analysis Center/California Institute of Technology, funded by the National Aeronautics and Space Administration and the National Science Foundation. This work is supported by the Science and Engineering Research Board (SERB), a statutory body under Department of Science and Technology (DST), Government of India, under Fast Track scheme for Young Scientist (SR/FTP/PS-092/2011).



\end{document}